\documentclass[%
reprint,
superscriptaddress,
 amsmath,amssymb,
 aps,
prb,
]{revtex4-2}
\usepackage[english]{babel}
\makeatletter
\let\ORIbbl@fixname\bbl@fixname
\def\bbl@fixname#1{%
	\@ifundefined{languagealias@\expandafter\string#1}
	{\ORIbbl@fixname#1}
	{\edef\languagename{\@nameuse{languagealias@#1}}}%
}
\newcommand{\definelanguagealias}[2]{%
	\@namedef{languagealias@#1}{#2}%
}
\makeatother
\definelanguagealias{en}{english}

\usepackage{graphicx}
\graphicspath{ {Images/} }
\usepackage{dcolumn}
\usepackage{bm}
\usepackage{amsmath}
\usepackage{mathtools}
\usepackage{physics}
\usepackage{hyperref}
\hypersetup{
	colorlinks=true,
	allcolors=blue
}
\usepackage{braket}

\newcommand{\xibf}{\boldsymbol{\xi}}
\newcommand{\kbf}{\mathbf{k}}

\newcommand{\Ebf}{\mathbf{E}}

\newcommand{\Abf}{\mathbf{A}}

\newcommand{\Omegabf}{\mathbf{\Omega}}
\newcommand{\rbf}{\mathbf{r}}

\newcommand{\pbf}{\mathbf{p}}

\begin{document}
\title{Microscopic analysis of high harmonic generation in semiconductors with degenerate bands}

\author{\surname{Le} Huu Thong}
\affiliation{Ho Chi Minh City Institute of Physics, Vietnam Academy of Science and Technology,  1 Mac Dinh Chi Street, District 1, Ho Chi Minh City, Vietnam}
\affiliation{Department of Physics, University of Science, Vietnam National University Ho Chi Minh City, 227 Nguyen Van Cu Street, District 5, Ho Chi Minh City, Vietnam}

\author{Cong Ngo}
\affiliation{Department of Physics and CeOPP, Universit\"at Paderborn, Warburger Strasse 100, D-33098 Paderborn, Germany}

\author{\surname{Huynh} Thanh Duc}
\affiliation{Ho Chi Minh City Institute of Physics, Vietnam Academy of Science and Technology,  1 Mac Dinh Chi Street, District 1, Ho Chi Minh City, Vietnam}

\author{Xiaohong Song}
\affiliation{Research Center for Advanced Optics and Photoelectronics, Department of Physics, College of Science, Shantou University, Shantou, Guangdong 515063, China}

\author{Torsten Meier}
\affiliation{Department of Physics and CeOPP, Universit\"at Paderborn, Warburger Strasse 100, D-33098 Paderborn, Germany}

\date{\today}
\begin{abstract}
Based on the multiband semiconductor Bloch equations a microscopic approach to high-harmonic generation in crystalline solids which is able to properly describe degenerate bands and band crossings is presented and analyzed.
It is well-known that numerical band structure calculations typically provide electronic wave functions with an undetermined k-dependent phase which results in matrix elements which contain arbitrary k-dependent phases.
In addition, such approaches usually mix degenerate bands and bands with an energy difference smaller than the numerical precision in an arbitrary way for each point in k-space.
These ambiguities are problematic if one considers the dynamics induced by electric fields since the matrix elements of the position operator involve a derivative of the wave functions with respect to k.
When the light-matter interaction is described in the length gauge, the problem of arbitrary phases and degenerate subspace mixing of Bloch states is solved by adopting a smooth gauge transformation along the field direction. The results obtained within this method are validated by comparing with calculations in the velocity gauge.
Although we obtain in both gauges the same overall result, the length gauge is advantageous since it converges with a smaller number of bands and thus requires significantly less numerical effort than the velocity gauge. Also an unique distinction between inter- and intraband contributions and thus an instructive physical interpretation is possible in the length gauge whereas in the velocity gauge this is unclear. The computed polarization-direction-dependent high-harmonic spectra agree well with experimental data reported for GaAs. Furthermore, it is demonstrated that, under proper conditions, the Berry curvature is largely responsible for the even-order harmonics which are polarized perpendicular to the driving field.
\end{abstract}

\maketitle

\section{Introduction}
	Since its discovery in 2010 \cite{Ghimire2011}, high-harmonic generation (HHG) from solids has been extensively studied. Under the excitation by an intense laser field, solid crystals can emit high-order harmonics of the driving frequency, over a very broad spectrum and with promising properties for applications. While the role of different mechanisms are still being discussed, the semiconductor Bloch equations (SBE) serve as a generic versatile approach that yields good agreement with several experiments and allows for analyzing fine details of the generation process and the dynamics of the photoexcitations \cite{Golde2008,Golde2011, Luu2016, Vampa2014, Schubert2014, Hohenleutner2015,Song2019}.

The interaction between matter and light fields is usually described in either the length gauge (LG) or the velocity gauge (VG). Early models for HHG in solids were implemented exclusively
in the LG. In this gauge, the HHG spectrum is usually analyzed in terms of interband and intraband components, which are believed to originate from optical transition between different bands and the electrons' acceleration within each band,  respectively. Such an intuitive picture has been shown to be reliable in the LG, it is, however, questionable for the VG \cite{Ernotte2018}.

However, in the LG the SBE have the disadvantage that they requires the Bloch-state basis to be smooth in reciprocal space (k-space). When the Bloch states at each k-point are obtained from numerical diagonalization they are prone to suffer from random phases and swapping of the band index of degenerate states. This results in phase jumps of the complex transition dipole $ \xibf(\kbf) $ between neighboring $\mathbf{k}$ vectors.  A primitive remedy to this problem was to only consider the absolute value $\abs{\xibf(\kbf)} $, which is an often used approximation when solving the SBE \cite{Vampa2014}. Recently, the transition-dipole's phase has been claimed to carry information on the crystal's symmetry and to be essential for the description of even-order harmonics \cite{Jiang2018}. To incorporate the transition-dipole phases properly when solving the SBE, it is required to implement a gauge which results in Bloch functions which vary smoothly in k-space. For non-degenerate energy bands, methods to fix the phases have been proposed and applied to investigate HHG \cite{Lindefelt2004,Wu2015,Li2019,Yue2020}.
In the presence of degenerate and crossing bands, this issue has, however, rarely been addressed, since the  Bloch states are further subject to an arbitrary unitary transformation  within each degenerate subspace. Here we apply the method proposed by Virk and Sipe \cite{Virk2007} to construct a smooth parallel-transport gauge of degenerate Bloch functions which allows us to solve the SBE in the LG and compute HHG emission spectra.

Several recent publications \cite{Foldi2017, Wismer2018, Yue2020} have demonstrated that HHG calculations in both the VG and the LG agree if in both gauges the relevant band are included and no further approximations, in particular, no approximations that are not gauge invariant are made. Here, we confirm the gauge independence of our results which verifies the correctness of our gauge procedure.
Even though we obtain in both gauges the same overall results, the LG converges with a significantly smaller number of bands (less than half) than required in the VG and thus the numerical effort of the calculations is strongly reduced. Furthermore, unlike in the VG, in the LG a unique distinction between inter- and intraband contributions and thus an instructive physical interpretation is feasible and physically sound.

On the other hand, in HHG experiments it has been reported that there exist components whose polarizations are perpendicular to that of a linearly-polarized driving electric field. This emission has been ascribed to originate from the crystal's Berry curvature via the anomalous-velocity formula within a semi-classical approximation \cite{Liu2017, Luu2018}.
For much weaker THz fields that follow an optical excitation with circularly-polarized light, similar Berry-curvature-induced so called \textit{anomalous currents} that flow in a perpendicular direction have been proposed and measured in semiconductor quantum wells \cite{VS2011,Bieler2015}.
In this work, we study that \textit{perpendicular HHG component} within the microscopic theory of the SBE. We find that the Berry curvature indeed contributes predominantly to the perpendicular emission calculated within the microscopic theory, though only when the laser's frequency is small compared to the band-gap and the field strength is strong enough to drive electrons to regions of large Berry curvature.

The paper is organized as follows. In Section~\ref{sec.SBE} we present the SBE  in both the LG and the VG. Section~\ref{sec.ParallelTransport} describes the method of constructing a proper parallel-transport gauge for linearly polarized incident fields including degenerate Bloch states. Numerical results are presented and discussed in Section~\ref{sec.Results} where we consider GaAs whose band structure and matrix elements are obtained from the 30-band kp method \cite{Richard2004}. In the results section we focus on three aspects: the equivalence between calculations in the LG and in the VG, the comparison with a reported HHG experiment performed on GaAs \cite{Xia2018} and the contribution of the Berry curvature to perpendicularly polarized HHG emission.
We close with a brief summary presented in Section~\ref{concl}.

\section{Theoretical methods}\label{sec.Methods}
In this section we first present the derivation of the SBE in LG and VG and then describe the
parallel-transport gauge which is applied to be able to numerically solve the SBE in LG.

\subsection{The SBEs in LG and VG}\label{sec.SBE}
In second-quantization the Hamiltonian describing Bloch electrons interacting with a light field is written in LG as
\begin{eqnarray}
H^\mathrm{}&=&H_0
- e\mathbf{E}(t)\cdot \sum_{\lambda\lambda'\mathbf{k}\mathbf{k'}}\mathbf{r}_{\lambda\lambda'}(\mathbf{k},\mathbf{k'})a^\dagger_{\lambda\mathbf{k}}a^{}_{\lambda'\mathbf{k'}},
\end{eqnarray}
where $H_0$ is the Hamiltonian of the crystal, $a^\dagger_{\lambda\mathbf{k}}$ $(a^{}_{\lambda\mathbf{k}})$ is the creation (annihilation) operator of an electron with wave vector $\mathbf{k}$ in band $\lambda$, $\mathbf{E}(t)$ is the electric field (whose spatial dependence is neglected here in the long-wavelength limit), and $\mathbf{r}$ is the position operator.
Although its explicit form is not required for the following derivations, we would like to mention that in our numerical evaluations we take $H_0$ as the single-particle Hamiltonian of the Bloch electrons, i.e., $H_0 = \sum_{\lambda \mathbf{k}}  \epsilon_\lambda (\mathbf{k}) a^\dagger_{\lambda\mathbf{k}}a^{}_{\lambda\mathbf{k}}$, where $\epsilon_\lambda (\mathbf{k})$ is the band structure.

In the Bloch basis, the position operator can be represented as \cite{Blount1962}
	\begin{eqnarray}\label{eq.PositionOperator}
		\rbf_{\lambda \lambda'}(\kbf ,\kbf')
		=  \left(i\delta_{\lambda\lambda'} \nabla_{\kbf} + \xibf_{\lambda\lambda'}(\kbf)\right)\delta({\kbf-\kbf'}),
	\end{eqnarray}
where $ \xibf_{\lambda\lambda'}(\kbf) =-i
	\braket{\nabla_{\kbf}u_{\lambda \kbf}|u_{\lambda' \kbf}} $ is the transition dipole matrix element and $ \ket{u_{\lambda \kbf}}$ is the periodic part of the Bloch function. While one has the gauge freedom in choosing the phase of $ \ket{u_{\lambda \kbf}}$, this representation of $ \rbf $ is limited to gauges that smoothen the $k$-dependence of the wave functions, such that the above derivative with respect to
$k$ is well-defined.

The semiconductor Bloch equations \cite{Haug2004}, that describe the dynamical optoelectronic response, can be expressed in terms of the reduced density matrix elements  $ \rho_{\lambda\lambda'}(\kbf) = \braket{a^\dagger_{\lambda' \kbf} a_{\lambda \kbf}} $ as
	\begin{eqnarray}
		\frac{d\rho(\kbf)}{dt}
		&=& -\frac{i}{\hbar} \comm{H_0(\kbf) -e\Ebf(t)\cdot\xibf(\kbf)}{\rho(\kbf)} \nonumber\\ &&-\frac{e}{\hbar}\Ebf(t)\cdot\nabla_\kbf \rho(\kbf), \label{eq.SBE_LG}
	\end{eqnarray}
where $ H_0(\kbf) = e^{-i\kbf\cdot\rbf} H_0 e^{i\kbf\cdot\rbf} $.
One can see that the equations of motion at different $\kbf$ vectors are coupled  by the derivative term $ \nabla_{\kbf} \rho(\kbf) $, which stem from the position operator in Eq.~\eqref{eq.PositionOperator}.
It is this term that poses the requirement of constructing a smooth gauge of the Bloch functions in order to be able to solve the SBE.
In many previous works this problem has been ignored and it has been assumed that the transition dipoles are independent of $k$ or have a simple $k$-dependence
or often only their absolute values were considered.
The proper smooth gauge is, however, unavoidable if one wants to work in the preferred LG and consider the full information in the form of complex matrix elements which arise from band structure calculations like $\mathbf{k}\cdot\mathbf{p}$ or density functional theory that contain arbitrary $k$-dependent phases arising from the numerical diagonalization.

In the VG the SBE read
	\begin{eqnarray}
		&&	\frac{d\rho(\kbf)}{dt} =  -\frac{i}{\hbar} \comm{H_0(\kbf) - \frac{e}{m} \mathbf{A}(t)\cdot\pbf(\kbf)}{\rho(\kbf)},\label{eq.SBE_VG}
	\end{eqnarray}
	 where $ \Abf(t) = -\int_{-\infty}^t \Ebf(t')dt' $ is the vector potential and $ \pbf(\kbf) = \dfrac{m}{\hbar} \nabla_\kbf H_0(\kbf) $ is the canonical momentum matrix.
In contrast to Eq.~(\ref{eq.SBE_LG}), the equations for the reduced density matrix in VG, Eq.~(\ref{eq.SBE_VG}), at different $\mathbf{k}$ vectors are independent of each other, hence a gauge choice ensuring the smoothness of the wave function in $k$-space is not necessary.
As is shown below, whereas we do obtain very similar final results in both gauges the LG is more efficient since a smaller number of bands is required to obtain converged results and, in addition, the interpretation in terms inter- and intraband contributions seems to be more adequate.
The transformation between the LG and the VG can be performed by the unitary operator $ Q = e^{-ie\Abf(t)\cdot\rbf/\hbar} $\cite{Foldi2017,Yue2020}.

It is noted that the common expression for the decoherence term, which describes the exponential decay of off-diagonal elements of the density matrix,
	\begin{equation}
		\left(\frac{d \rho(\kbf)}{d t} \Big\rvert_{\text{decoh}}\right)_{\lambda\lambda'}= -\frac{\rho_{\lambda\lambda'}(\kbf)}{T_2} (1 - \delta_{\lambda\lambda'})
		\label{eq.OldT2},
	\end{equation}
is not gauge invariant and the gauge transformation of this term can only be done by numerical calculations \cite{Yue2020}.
Instead of Eq.~(\ref{eq.OldT2}), we therefore use its gauge-covariant form that was proposed in Ref.~\citenum{Wismer2018} and allows for a simple transformation between these two gauges. The decoherence term is given in LG and VG, respectively, by \cite{Wismer2018}
	\begin{eqnarray}\label{eq.NewT2LG}
		\left.\frac{d\rho(\kbf)}{dt} \right|_{\text{decoh}} &=&
-\frac{1}{T_2 E_g^2} \left[H_0(\kbf),\left[ H_0(\kbf),\rho(\kbf)\right]\right]
	\end{eqnarray}
and
	\begin{eqnarray}\label{eq.NewT2VG}
		\left.\frac{d\rho(\kbf)}{dt} \right|_{\text{decoh}} &=&
-\frac{1}{T_2 E_g^2} \left[H_0(\kbf)-\frac{e}{m} \mathbf{A}(t)\cdot\pbf(\kbf),\right.\nonumber\\
&&	\left.\left[H_0(\kbf) -\frac{e}{m} \mathbf{A}(t)\cdot\pbf(\kbf),\rho(\kbf)\right]\right],
	\end{eqnarray}
where $T_2$ is the phenomenological decoherence time and $E_g$ is the bandgap energy.

\subsection{The parallel-transport gauge}\label{sec.ParallelTransport}

In order to solve the the SBE in LG, we follow Virk and Sipe \cite{Virk2007} and construct a local gauge transformation of the Bloch functions. The procedure is based on the $\mathbf{k}\cdot\mathbf{p}$ perturbation theory for the band structure.
To first order in $\Delta\kbf$, the perturbation theory provides the relation between the eigenstates at $\kbf$ and $\kbf+\Delta\kbf$ \cite{Virk2007,Lax2001}
	\begin{equation}
	\ket{u_{\lambda\kbf+\Delta\kbf}} =\sum_{\nu\mu} \ket{ u_{\mu\kbf}} (\delta_{\nu\mu} - i\Delta\kbf\cdot \xibf^\text{ter}_{\mu\nu}(\kbf)) g_{\nu\lambda}({\kbf, \kbf+\Delta\kbf}),
	\end{equation}
	where $ g(\kbf, \kbf+\Delta\kbf)=\exp(-i\Delta\kbf\cdot\xibf^\text{tra}(\kbf)) $ is an unitary matrix accounting for arbitrary phase factors. The transition dipole matrix $\xibf(\kbf)$ is separated into $\xibf^\text{tra}(\kbf)$ and $\xibf^\text{ter}(\kbf)$ which contains matrix elements between connected states and between disconnected states, respectively. Here, two states $|u_{\lambda\kbf}\rangle$ and $|u_{\nu\kbf}\rangle$ are called connected if they have the same energy at at least one $k$-point in the Brillouin zone (BZ), otherwise they are called disconnected.

Introducing the overlap matrices between the two bases at different $k$-points $ S^\text{tra}_{\lambda\lambda'}({\kbf,\kbf'}) = \braket{u_{\lambda\kbf}| u_{\lambda'\kbf'}} \Delta_{\lambda\lambda'} $ and $S^\text{ter}_{\lambda\lambda'}({\kbf,\kbf'}) = \braket{u_{\lambda\kbf}| u_{\lambda'\kbf'}} \left(1-\Delta_{\lambda\lambda'}\right) $, where $\Delta_{\lambda\mu}=1$ if $\lambda=\mu$ or if two bands $\lambda$ and $\mu$ are connected and $\Delta_{\lambda\mu}=0$ otherwise, the above perturbative relation gives
  \begin{equation}\label{gtra}
		g(\kbf,\kbf+\Delta\kbf) =  S^\text{tra}({\kbf,\kbf+\Delta\kbf})
	\end{equation}
and
	\begin{equation}\label{xiter}
                -  i\Delta\kbf \cdot \xibf^\text{ter}(\kbf) g(\kbf,\kbf+\Delta\kbf)=S^\text{ter}(\kbf,\kbf+\Delta\kbf).
	\end{equation}
Eqs.~(\ref{gtra}) and (\ref{xiter}) allow one to calculate $g(\kbf,\kbf+\Delta\kbf)$ and $\xibf^\text{ter}(\kbf)$ from the overlap matrices which are determined by the wave functions obtained from the band structure calculations. To eliminate the relative phase of eigenstates between $\kbf$ and $\kbf+\Delta\kbf$ one performs the gauge transformation
\begin{eqnarray}\label{gaugetran}
\ket{u_{\lambda\kbf+\Delta\kbf}}\longmapsto \sum_\mu g^*_{\lambda\mu}(\kbf,\kbf+\Delta\kbf)\ket{u_{\mu\kbf+\Delta\kbf}}.
\end{eqnarray}
Although in practical calculations $g(\kbf,\kbf+\Delta\kbf)$ is not exactly unitary, it can be made unitary by a singular value decomposition (SVD) algorithm.

By extending Eq.~(\ref{gaugetran}) to a series of $k$-points on a straight line in $j$-direction, with unit vector $\hat{\mathbf{e}}_j$, and starting from the first point $\kbf_0$, one constructs a gauge transformation matrix
\begin{eqnarray}
W^j(\kbf_0,\kbf)=g(\kbf_0,\kbf_0+\hat{\bf{e}}_j\Delta k)...g(\kbf-\hat{\bf{e}}_j\Delta k,\kbf)
\end{eqnarray}
that removes the arbitrary relative phase of the eigenstates between every two nearest neighboring $k$-points, which makes the Bloch functions smooth and differentiable with respect to $\kbf$
	\begin{eqnarray}
		\ket{u_{\lambda\kbf}} \longmapsto  \sum_\mu W^{j*}_{\lambda\mu}(\kbf_0,\kbf) \ket{ u_{\mu\kbf}}.
	\end{eqnarray}
Mathematically, the above method is equivalent to the optimal alignment procedure presented in Ref.~\citenum{Vanderbilt2018}, section 3.6.
	
In the new basis of Bloch functions, the projection onto the $j$-direction of $\xibf^\text{tra}(\kbf)$ vanishes and the projection of $\xibf^\text{ter}(\kbf)$ given by
	\begin{eqnarray}\label{Retrieve}
		&&\xi^\text{ter}_j({\kbf}) = \frac{i}{2\Delta k}W^j({\kbf_0,\kbf})\nonumber\\
		&&\ \ \ \times \Big[S^\text{ter}(\kbf,\kbf+\hat{\mathbf{e}}_j\Delta k) S^{\text{tra}\dagger}({\kbf,\kbf+\hat{\mathbf{e}}_j\Delta k}) \\
		&&\ \ \ -S^{\text{ter}\dagger}({\kbf-\hat{\mathbf{e}}_j\Delta k,\kbf}) S^\text{tra}({\kbf-\hat{\mathbf{e}}_j\Delta k,\kbf})\Big] W^{j\dagger}({\kbf_0,\kbf}) \nonumber
	\end{eqnarray}
is a smooth function of $\kbf$. The above gauge of the Bloch functions is called the parallel-transport gauge \cite{Xiao2010}.

\section{HHG from G\lowercase{a}A\lowercase{s}: Results and discussion}\label{sec.Results}

By integrating the SBE in LG (\ref{eq.SBE_LG}) or VG (\ref{eq.SBE_VG}) for an one-dimensional $k$-grid that is parallel to the polarization-direction of the electric field, we obtain the dynamics of the density matrix $\rho(k,t)$ which determines the time-dependent electric current density via
\begin{eqnarray}\label{current}
\mathbf{J}(t)=e\sum_{\lambda\lambda' k}\mathbf{v}_{\lambda\lambda'}(k)\rho_{\lambda'\lambda}(k,t),
\end{eqnarray}
where $\mathbf{v}=\frac{i}{\hbar}\left[H,\mathbf{r}\right]$ is the velocity operator. The momentum space representations of $\mathbf{v}$ are $\mathbf{v}(k)=\mathbf{p}(k)/m$ in LG and $\mathbf{v}(k)=\left(\mathbf{p}(k)-e\mathbf{A}(t)\right)/m$ in VG.

The photoexcited current oscillates rapidly and contains multiples of the excitation frequency which correspond to high-harmonic radiation. From the Fourier transform of current density we obtain the spectrum of the emission intensity
\begin{eqnarray}
I_{\text{HHG}}(\omega)\propto \abs{J(\omega)}^2.
\end{eqnarray}

In the following we carry out numerical calculations for bulk GaAs. The electronic band structure and wave functions of GaAs for the entire Brillouin zone (BZ) are obtained from a 30-band $\mathbf{k}\cdot\mathbf{p}$ model \cite{Richard2004}. This quite sophisticated model includes spin-orbit coupling and allows to describe the inversion asymmetry of GaAs crystal.

The electric field of the exciting laser pulse is described by
\begin{eqnarray}
\mathbf{E}(t)=E_0\hat{\mathbf{e}}\ e^{-2\ln(2)t^2/\tau^2}\sin(\omega_0 t),
\end{eqnarray}
where $\hat{\mathbf{e}}$ denotes the polarization direction, $E_0$ is the maximal amplitude, $\tau$ is the pulse duration (FWHM of the Gaussian envelope), and $\omega_0$ is the central light frequency.

\subsection{Comparison between LG and VG}\label{sec.LGVG}

\begin{figure}[t]
		\centering
		\includegraphics[]{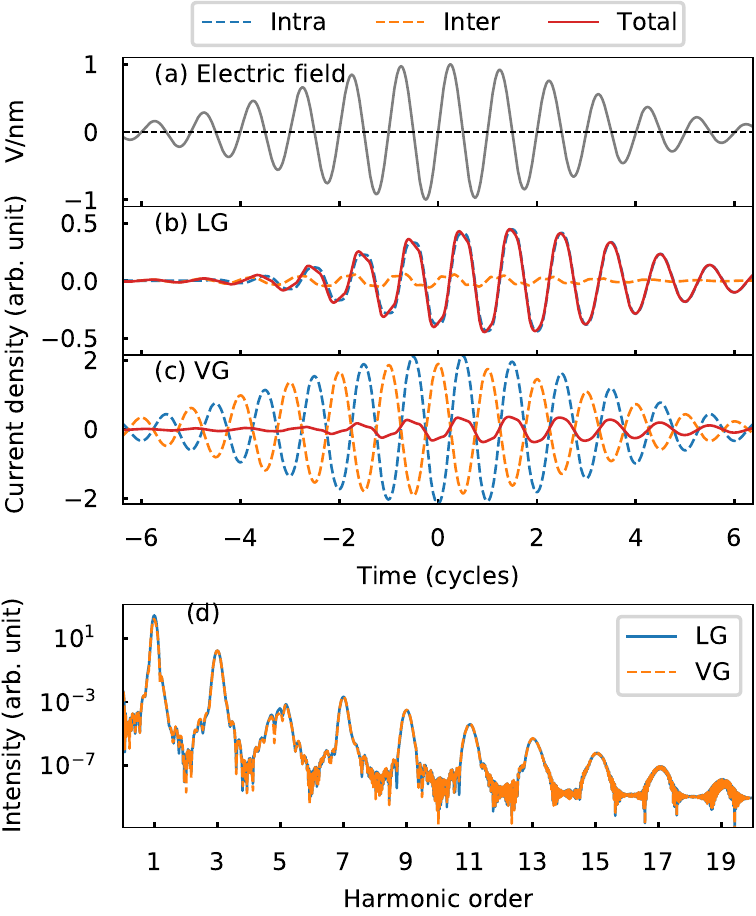}
		\caption{(a) shows the electric field of the time-dependent exciting THz laser pulse.
		(b) and (c) show the temporal dynamics of the current density calculated for a THz field that is linearly polarized in the $[100]$ crystallographic direction (corresponding to the $\Gamma \text{X}$ direction in the BZ) in the LG and the VG, respectively. The total current (red solid lines) is given by the sum of the intraband (blue dashed lines) and the interband (orange dashed lines) currents.
		(c) Intensity spectra of the emitted high harmonics in the two gauges.}
		\label{fig.CompareLGVG}
	\end{figure}

The computed time-dependent photocurrent for the two gauges and the resulting emission spectra are presented in Fig.~\ref{fig.CompareLGVG}. In the numerical simulations, the driving laser is linearly polarized in the $[100]$ crystallographic direction ($\Gamma \text{X}$ direction in the BZ) and has an amplitude of $E_0=10~\text{MV/cm}$, $\tau= 60~\text{fs}$, and a center frequency of $\hbar\omega_0=0.38~\text{eV}$.

Fig.~\ref{fig.CompareLGVG}(b) shows the dynamics of the current density calculated in LG. The red-solid line represents the total current, while the blue and orange dashed lines display its two components corresponding to the intraband (diagonal) and interband (off-diagonal) contributions, respectively. While the interband current is in phase with the electric field (grey line in Fig.~\ref{fig.CompareLGVG}(a)), the intraband current lags behind the field by a phase of $\pi/2$. This phase lag of the intraband current can be explained by the acceleration theorem for the electron wave packet motion $\hbar\dot{\mathbf{k}}=e\mathbf{E}(t)$, i.e., the change of $\mathbf{k}$ of proportional to the time integral over $\mathbf{E}(t)$.

Fig.~\ref{fig.CompareLGVG}(c) is the same as Fig.~\ref{fig.CompareLGVG}(b) but calculated in the VG. Compared to the LG, the interband and intraband currents in the VG are both much stronger
 and are apart from their opposite sign  very similar to each other. 
 Obviously, these currents are gauge dependent and have therefore in the VG no clear physical significance \cite{Foldi2017,Ernotte2018}. Due to the strong cancellations between the interband and intraband currents in the VG, the resulting gauge-independent total current is identical to that obtained in the LG.

We would like to emphasize that the calculation in the LG is already converged when the 6 highest valence and the 8 lowest conduction bands are considered in the SBE, while the VG requires that all 30 bands included. Since in the VG the subcurrent components are gauge dependent, a full sum (over all available bands) is necessary so that their gauge dependencies basically compensate each other which results in a total current which is independent of the chosen gauge (if sufficiently many bands can be considered). Since the effort for numerical solutions of the SBE (without including many-body interactions) scales quadratically with the number of bands, the evaluations in the LG are significantly faster than those in the VG and in addition in the VG and also the required computer memory is reduced similarly.

In Fig.~\ref{fig.CompareLGVG}(c) we show the spectral intensity of the emitted HHG computed in LG and VG. Since the incident laser pulse is polarized in the $\Gamma\text{X}$ direction, the spectrum consists of only odd-order harmonics.  Just like the time-dependent total currents, the HHG spectra in the two gauges are almost identical, i.e., the 30 band in our band-structure model are sufficient to ensure convergence of the VG results for the considered excitation conditions.

	\subsection{Comparison to experiment}\label{sec.Experiment}

Next, we compare our numerical results with experimental data reported in Ref.~\citenum{Xia2018}. In our simulations, we consider the experimental conditions and use the following parameters for the driving laser: a photon energy of $\hbar\omega_0=E_g/4$, a pulse duration (FWHM) of 5 laser cycles, and a maximal electric field  amplitude of $E_0=10~\text{MV/cm}$.  The only fitting parameter is the decoherence time $T_2 $, which is taken to be on the order of a few femtosecond as in several previous HHG studies \cite{Luu2016}.
As shown recently, such very small dephasing times can be justified as they yield similar HHG spectra as arising when including propagation effects  \cite{Kilen2020}.
Both types of the decoherence term described in Sec.~\ref{sec.SBE} can be used to match the experiment with nearly the same value of $T_2 $. Here, we show the calculated results in the LG using the decoherence term Eq.~\eqref{eq.NewT2LG} with $T_2 = 3$~fs (which corresponds to 3/10 of the laser period).

	\begin{figure}[t]
		\centering
		\includegraphics[]{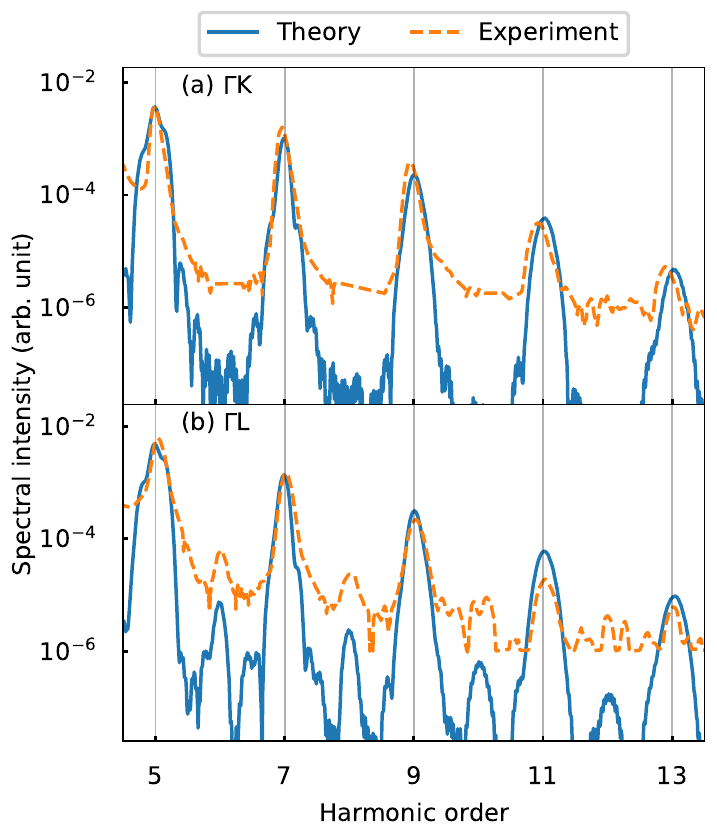}
		\caption{Calculated HHG spectra of GaAs with the exciting laser field polarized linearly along the (a) $\Gamma\text{K}$ and (b) $\Gamma\text{L}$ directions, respectively, in comparison to experimental data which is extracted from Ref.~\citenum{Xia2018}.}
		\label{fig.Experiment}
	\end{figure}

For the case that the incident light is polarized linearly in the $\Gamma\text{K}$ direction, both the calculated and the measured HHG spectra exhibit only odd-order harmonic peaks, see Fig.~\ref{fig.Experiment}(a), since the inversion asymmetry of GaAs is not probed in this excitation configuration. Our theoretical simulations describe the decrease of the intensity of the odd-order harmonics with increasing harmonic order in very good agreement with the experimental observations. An important point is that an optical excitation polarized in the $\Gamma\text{K}$ direction does also produce harmonics that are polarized perpendicularly to the direction of the incident laser field. This perpendicular component of HHG was, however, not measured in \cite{Xia2018}. We will present our theoretical investigation of this component in the next section.

When the incident light is polarized linearly in the $\Gamma\text{L}$ direction ($[111]$ crystallographic direction), besides the odd-order harmonics, also even-order harmonics with somewhat weaker intensities arise, see Fig.~\ref{fig.Experiment}(b). The even-order harmonics originate from the inversion asymmetry of GaAs crystal. In the $30$-band $\mathbf{k}\cdot \mathbf{p}$ model \cite{Richard2004}, $P'$ is the only parameter that models this asymmetry and consequently the even-order harmonics disappear if one artificially sets $P'=0$. Compared to experiment, the calculated even-order harmonics are somewhat weaker which is probably the case because the 30-band $\mathbf{k}\cdot\mathbf{p}$ model underestimates the inversion asymmetry of GaAs as it uses a smaller $P'$ parameter than other $\mathbf{k}\cdot\mathbf{p}$ models \cite{Mayer1991,Saidi2010}, e.g., in the 14-band $\mathbf{k}\cdot\mathbf{p}$ model this term is almost ten times larger. For example, if we manually double the value of $P'$ in the 30-band model (not shown in figure),
the amplitude of the even-order harmonics increases accordingly and the calculated even-order peaks
agree better with experiment. This is one example showing that HHG spectra can be used to gain information on the atomic structure as a detailed comparison between measurements and calculations 
can be used to improve the band structure models \cite{Tancogne2017,Lanin2017}.
	
	\subsection{Anomalous perpendicular currents arising from Berry curvature} \label{sec.Berry}

Although we solve the SBE here in just one $k$-space dimension, the current density obtained from $\eqref{current}$ is still a three-dimensional quantity. For example, when the laser field is polarized in the $[110]$ ($\Gamma\text{K}$) direction, beside the main current that flows parallel to the field direction, the photoexcited current also has a smaller perpendicular component that flows in the $[001]$ ($\Gamma\text{X}$) direction.
We evaluate this perpendicular current component in two ways: directly from the microscopic approach using Eq.~\eqref{current} and indirectly from the perturbative equation for the anomalous velocity (Berry curvature approach)
	\begin{equation}
		\mathbf{J}_\perp^{a}(t) = - \frac{e^2}{\hbar}\Ebf(t) \times \expval{\Omegabf},\label{eq.Anomalous}
	\end{equation}
where $\expval{\Omegabf} = \sum\limits_{\lambda\lambda'k} \Omegabf_{\lambda\lambda'}(k) \rho_{\lambda'\lambda}(k,t) $ is the average of the Berry curvature
	\begin{equation}
	\Omegabf_{\lambda\lambda'}(k)
	= i\sum_{\mu} \xibf^\text{ter}_{\lambda\mu}(k) \times \xibf^\text{ter}_{\mu\lambda'}(k).
	\end{equation}

	\begin{figure}[t]
		\centering
		\includegraphics[]{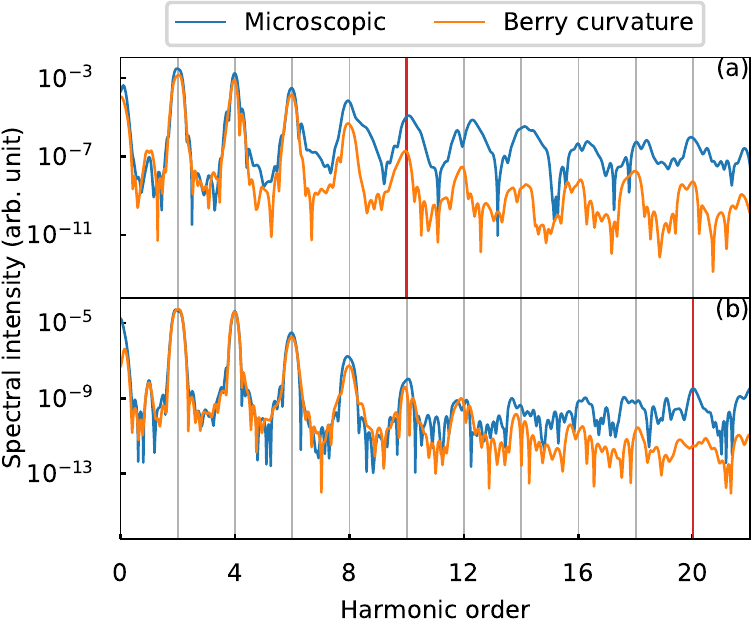}
		\caption{Spectra of the HHG radiation with perpendicular polarization calculated by the microscopic and the Berry curvature approaches for an excitation photon energy of (a) $ \hbar\omega_0 = E_g/10 $ and (b) $ \hbar\omega_0 = E_g/20 $. The vertical red lines indicate the bandgap energy $E_g$.
The amplitudes of the THz fields are $E_0=8$~MV/cm in (a) and  $E_0=4$~MV/cm in (b) such that the
ratio $eE_0/\hbar\omega_0$ remains the same.}
		\label{fig.Berry}
	\end{figure}
	
	We solve the SBE in the LG using the decoherence term Eq.~\eqref{eq.OldT2} with $T_2=1/5$ of the laser period. The perpendicularly polarized HHG spectra calculated by the microscopic (blue line) and Berry curvature (orange line) approaches are shown in Fig.~\ref{fig.Berry}(a) for a photon energy of $\hbar\omega_0=E_g/10$. We find that the perpendicularly-polarized HHG component is dominated by even-order harmonics and its intensity is about two orders smaller than that of the parallel HHG component. The overall good agreement between two approaches confirms the Berry curvature as the origin of the perpendicular current component. Since $\Omegabf(k)$ is an odd function of $k$, the anomalous velocity has a fundamental frequency of $2\omega_0$, and hence the perpendicular HHG basically contains only even-order harmonics \cite{Liu2017,Luu2018}.

The results shown in Fig.~\ref{fig.Berry}(b) are obtained for a more slowly varying laser with $\hbar\omega_0=E_g/20$. To ensure that the region which the electronic wavepacket traverses to is the same as in the previous case, we adjust the field amplitude so that $eE_0/\hbar\omega_0$ is unchanged. We note that the expression for anomalous velocity, Eq.~\eqref{eq.Anomalous}, is derived from the first-order adiabatic perturbation theory. Thus, the difference between two approaches describes the contributions beyond this approximation. Because the adiabatic condition holds better for smaller laser frequencies, the agreement between two approaches in Fig.~\ref{fig.Berry}(b) is closer than that in Fig.~\ref{fig.Berry}(a).

An intrinsic limitation of one-dimensional models is that they only allow to compute the interband contribution to the perpendicularly-polarized current but not the intraband one. Our result therefore do not include the perpendicular odd-order harmonics which were shown to be produced predominantly by the intraband current for fields with high amplitudes \cite{Kaneshima2018}.

\section{CONCLUSIONS}
\label{concl}
We present and analyze a microscopic approach to high-harmonic generation in solids with degenerate bands and crossing band. We confirm that the calculations in the LG and in the VG produce the same results if sufficiently many bands are included in the numerical solutions and gauge-dependent approximation, e.g., when including dephasing, are avoided.

To be able to solve the SBE in the LG we implement a parallel-transport gauge which is able to properly treat degenerate bands and results in wave functions and matrix elements which vary smoothly as function of k. To obtain converged HHG results in the LG a bit less than half the number of bands is required than in the VG. Thus the numerical effort is strongly reduced in the LG and, furthermore, the LG allows to distinguish between inter- and intraband contributions and thus provides an instructive physical interpretation.

Our calculated results for HHG from GaAs are in good agreement with recent experimental data on the polarization-direction dependence. Furthermore, our approach is able to describe the perpendicularly-polarized even-order harmonics. These are  caused by the Berry curvature and also have contributions beyond the adiabatic approximation that are included in our microscopic approach.

The approach introduced here is very general, can be combined with various band structure computation methods, and is applicable to other strong field phenomena also involving resonant optical fields.

\begin{acknowledgments}
This work is funded by the Vietnam National Foundation for Science and Technology
Development (NAFOSTED) under the grant No. 103.01-2017.42, the Deutsche Forschungsgemeinschaft (DFG) through project ME 1916/4,
the National Natural Science Foundation of China under the grant No. 12074240, and by the Sino-German Mobility Programme (Grant No. M-0031).
We thank the PC$^2$ (Paderborn Center for Parallel Computing) for providing computing time.
\end{acknowledgments}

\end{document}